\documentstyle[11pt,psfig]{article}
\begin{document}

\begin{flushright}
UMDGR-97-112\\
gr-qc/9706027\\
\end{flushright} 
\vskip 1cm

\begin{center}
 
{\Large \bf Degenerate Metric Phase Boundaries}
 
\vskip 5mm
{Ingemar Bengtsson\footnote{E-mail: ingemar@vana.physto.se} 
\\Fysikum, Stockholm University, Box 6730\\
S-113 85 Stockholm, Sweden} 
\\and\\
\vskip 5mm
{Ted Jacobson\footnote{E-mail: jacobson@physics.umd.edu}
\\Department of Physics, University of
Maryland\\ College Park, MD 20742-4111, USA\\} 
\end{center}
 
\vskip 5mm

\begin{abstract}
{The structure of boundaries between degenerate and nondegenerate
solutions of Ashtekar's canonical reformulation of Einstein's equations 
is studied. Several examples are given of such ``phase boundaries" in
which the metric is degenerate on one side of a null hypersurface and
non-degenerate on the other side. These include portions of  flat space,
Schwarzschild, and plane wave solutions joined to degenerate regions. In
the last  case, the wave collides with a planar phase boundary and
continues on with  the same curvature but degenerate triad, while the
phase boundary  continues in the opposite direction. We conjecture that
degenerate phase boundaries are always null.}
\end{abstract}

\section{Introduction}

The notion that the spacetime metric is a kind of ``order parameter"
$\langle g_{\mu\nu}\rangle\ne0$ has been suggested many times in 
many contexts. 
The peculiar fact that the metric is nonvanishing
in vacuum could be understood in this way. In string theory, 
where the metric is just a tiny part of a richer set of stringy 
degrees of freedom, such an interpretation appears particularly
natural. In fact it has been suggested\cite{AtickWitten} that above 
the Hagedorn temperature there is a transition to a ``topological"
phase with greater symmetry in which the metric vanishes. Degenerate 
metrics also occur in the loopy approach to canonical 
quantum Einstein gravity\cite{JacoSmol,Ashtekarbook}. 
In a quantum state whose wavefunction is based on holonomies of the
spin connection, the spatial metric has, microscopically, rank one on loops
away from loop intersections.

One way to explore the consequences of admitting degenerate metrics 
is to adopt a polynomial formulation of Einstein gravity. In fact
there are many ways to express Einstein gravity in polynomial
form, and indeed the possibility that degenerate metrics should
be admitted as {\em bona fide} solutions has been discussed many
times in the past (even by Einstein himself\cite{Einstein}). 
One reason
why this possibility has attracted little notice is that one must 
choose whether it is the covariant metric, the contravariant metric,
or some other related object that is to be allowed to become 
degenerate. The choice seems arbitrary, yet the resulting extensions
of the theory have quite different properties.

In this paper we adopt Ashtekar's Hamiltonian formulation\cite{Ashtekar}, 
which 
is polynomial in the canonical variables and so admits a degenerate
extension, and we study the structure of boundaries between 
nondegenerate and degenerate phases. Even within this category 
of degenerate extensions, there are inequivalent 
variations, and several investigations of these have already appeared
\cite{KoshDadi,Bengtsson,BombTorr,Vara,RomaJaco,Roma,Reisenberger,
Matschull,1+1,2+1,YonShiNak}. 
The variation we adopt is the original formulation
given by Ashtekar. It may turn out that this is not the ``right"
one. In fact, it has a negative energy problem\cite{Vara}
(which does not exist at least for one other degenerate extension
of GR\cite{Sam}), and some recent work suggests that the formulation
with scalar lapse and Hamiltonian constraint of unit weight 
might be more relevant for the quantum theory than  
Ashtekar's original formulation with a Hamiltonian constraint
of weight two\cite{Thiemann}. 
Nevertheless, it seems worthwhile to get a feeling
for the degenerate behavior of the various formulations, 
and a lot of work has already been done in quantum gravity 
using the formulation we are studying.

In Ashtekar's formulation one of the canonical variables is 
the spatial pull-back of the self-dual spin connection.
Its canonical conjugate $E_i^a$ is thus a triad of contravariant
spatial vector densities with tensor density weight one.
(Equivalently, their Levi-Civita duals are spatial two-forms.)
This triad is related to the spatial metric by

\begin{equation} E^a_iE^b_i = qq^{ab} \hspace{1cm} \Rightarrow 
\hspace{1cm} \det{E^a_i} = \det{q_{ab}} \equiv q \ . \label{1} 
\end{equation}

\noindent (Our notation is that $a, b, c, ...$ are spatial vector 
indices, while $i, j, k, ...$ are $SO(3)$ indices.) 
The phase space action has an elegant and polynomial form when 
expressed in these variables. In this paper we will study ``regular"
solutions to Ashtekar's Hamiltonian equations, that is, solutions in which
the canonical variables, shift vector, and lapse density all take finite 
values which, except for the lapse density, are allowed to vanish.  
Note that Ashtekar's lapse density $N$ is 
related to the usual scalar lapse M by 

\begin{equation} N = \frac{M}{\sqrt{q}} \ . \label{2} \end{equation}

\noindent If the spatial metric becomes degenerate, i.e. if 
$q \rightarrow 0$, then the requirement that $N$ should stay finite is a 
non-trivial requirement  
relating the coordinates and the covariant spacetime metric.
We will use both the canonical formalism and the 
covariant spacetime viewpoint to study the degenerate solutions.

The particular question that we wish to focus on
concerns the ``phase boundary" between two regions of spacetime having a 
degenerate metric on one side of the boundary. This question has been 
considered before by one of us \cite{Bengtsson}; the conclusion then was 
that the exterior Schwarzschild solution can be matched to a degenerate 
region across the event horizon. While this is true, we will see that it 
is also very misleading, since in fact the join can take place on much 
more general null surfaces. For example,  
a lightlike plane in Minkowski space can serve as such a phase boundary, 
and in the Schwarzschild solution one can join to a degenerate metric 
across any spherically symmetric null surface.

In section 2 of this paper we explain the general construction
for obtaining a solution with degenerate phase boundary by starting
with nondegenerate Ashtekar initial data, making a degenerate coordinate 
transformation, and evolving by Ashtekar's hamiltonian.
We then apply this method to flat space data to 
join Minkowski space to a degenerate region, and show how this 
solution can be generalized to include ``connection waves" \cite{1+1}
in the degenerate region. 

In section 3 we adopt a covariant approach to studying 
degenerate phase boundaries, imposing appropriate conditions
on the covariant metric that follow from the Ashtekar theory.
Using this method we find solutions in which portions of Minkowski and
Schwarzschild spacetimes are joined to degenerate regions.
We also study the stability of the flat space phase 
boundary, through the simple device of hitting it with a plane 
gravitational wave from the non-degenerate side. It 
turns out that the phase boundary remains null and the plane wave 
continues into the degenerate region where it becomes a degenerate
plane wave solution with the same curvature.  

In section 4 the question whether or not the phase
boundary must always be null is formulated and considered. 
We show that the field equations must play a role if this 
is to be the case, and we conjecture that it is related to the
fact that the characteristic surfaces of the nondegenerate theory are
null. Finally, we close in section 5 with a discussion of our results
and some open questions.

\section{Canonical approach}

\def\bz{\bar{z}}
\def\bA{\bar{A}}
\def\bE{\bar{E}}
\def\z{\zeta}
\def\be{\begin{equation}}
\def\ee{\end{equation}}

To begin with, let us show how any nondegenerate Ashtekar initial
data set $(E^a_i,A_a^i)$ 
can be modified to produce a solution with a degenerate
phase boundary at any initial spatial two-surface. 
This general construction was first discussed by
Varadarajan\cite{Vara}.  
Let the spatial coordinates be called $(x,y,\bz)$, and parametrise
$\bz$ by a function $\bz(z)$ in such a way that 
$\bz':=d\bz/dz\rightarrow0$
at $z=0$. Since the connection
$A_a^i$ is a covariant tensor, and the triad density $E^a_i$ 
is the Levi-Civita dual of a covariant tensor
$\Sigma_{bc\, i}=\epsilon_{abc}E^a_i$, the components of the
initial data remain finite in the new coordinate system
$(x,y,z)$. In fact, we have
\be
(A_x^i,\, A_y^i,\, A_z^i)= 
(\bA_x^i,\, \bA_y^i,\, \bz'\bA_z^i)
\ee
and
\be
(E^x_i,\, E^y_i,\, E^z_i)= 
(\bz'\bE^x_i,\, \bz'\bE^y_i,\, \bE^z_i).
\ee
Thus $A_z^i$, $E^x_i$, and $E^y_i$ go to zero at the surface
$z=0$. Now it is possible adopt this data for positive values of 
$z$, and to smoothly join it to data for negative values of $z$
in such a way that $A_z^i$, $E^x_i$, and $E^y_i$ remain
zero for all negative $z$. The initial value constraints will
of course be satisfied for positive $z$, since we have only
made a regular coordinate transformation there. For negative
$z$ the constraints require  
\begin{eqnarray}
\partial_z E^z_i&=&0\\
E^z_i\partial_z A_a^i&=&0,
\end{eqnarray}
as discussed in \cite{1+1}.
This new data has a ``phase boundary" between a nondegenerate
phase for $z>0$ and a degenerate phase for $z<0$. The data
can now be evolved using Ashtekar's hamiltonian, with finite
lapse and shift, to yield a regular solution to Ashtekar's
degenerate extension of general relativity. Let us see what this
procedure yields when performed starting with flat initial data.

Suppose the original ``barred" data is just $\bA_a^i=0$
and $\bE^a_i=\delta^a_i$. Then in the new coordinate
system we have $A_a^i=0$ and 
\begin{equation}
E^a_i={\rm diag}\Bigl(h(z),h(z),1\Bigr)
\label{data}
\end{equation}
where $h(z)=z'$ for $z>0$ and $h(z)=0$ for $z<0$.
This data satisfies all the constraints. Let us
evolve it according to the equations of motion 
with unit lapse and vanishing shift\cite{Ashtekarbook}:
\begin{eqnarray}
\partial_t A_a^i&=&i\epsilon_{ijk}E^b_jF_{ab}^k\\
\partial_t E^a_i&=&-i\epsilon_{ijk}E^b_jD_b E^a_k\label{Edot}
\end{eqnarray}
where $D_b$ denotes the covariant derivative with respect to the
connection $A_a^i$. Given our choice of initial data,
$A_a^i$ and $E^z_i$ are thus constant, equal to their
initial values. The remaining equations are best written
in terms of the complex linear combinations 
\begin{equation}
E^{a}_{\pm}:=E^a_1\pm iE^a_2,
\end{equation}
viz.
\be
\partial_t E^{a}_{\pm}=\pm\partial_z E^{a}_{\pm}.
\ee
The general solution is thus given by 
$E^{a}_{\pm}=E^{a}_{\pm}(z\pm t)$. Evaluating at time $t=0$ gives
\be
E^{x}_{\pm}=h(z\pm t),\qquad E^{y}_{\pm}=\pm ih(z\pm t).
\ee
The spacetime metric with unit lapse (density) and vanishing
shift is given in terms of the Ashtekar variables by 
\be
ds^2=-E dt^2 + q_{ab}dx^a dx^b,
\ee
where $E= {\rm det} E^a_i$ and $q_{ab}$ is the inverse of 
$q^{ab}=E^{-1}E^a_iE^{bi}$. In the present case we have
$E= h_+h_-$ and $q_{ab}={\rm diag}(1,1,h_+h_-)$,
where $h_\pm=h(z\pm t)$, and the metric becomes
\be
ds^2= h_+h_-( d\z^+d\z^-) + dx^2 + dy^2,
\ee 
where $\z^\pm=z\pm t$. Since one can integrate 
$h_\pm d\z^\pm=d\xi^\pm$, this is evidently just flat spacetime
as long as $h(z)$ is nowhere vanishing.

Now suppose we choose initial data of the form (\ref{data})
with $h(z)=0$ for $z<0$ and $h(z)>0$ for $z>0$. Then 
the solution is nondegenerate only in the right hand wedge $z>|t|$
(see figure). The boundary of this wedge is the ``phase
boundary". Minkowski spacetime has been smoothly spliced
onto a degenerate solution across this pair of intersecting
null surfaces. The triad takes the form:
\begin{equation}
E_i^a=\left(\begin{array}{ccc}h_+&h_-&0\\ih_+&-ih_-&0\\0&0&1\end{array}\right)
\end{equation}
where $h_\pm:=h(z\pm t)$, and the rows are $x,y,z$ components and the 
columns are $+$,$-$,$3$ components. In the left hand wedge both $h_+$ and
$h_-$ vanish, so both $E_i^a$ and $E_i^a E^{bi}$ have rank one. The triad
is therefore of type (1,1) in the Lewandowski-Wi\'{s}niewski
notation\cite{2+1}.
In the forward and backward wedges $h_-$ and $h_+$ vanish respectively,
so $E_i^a$ has rank two but $E_i^a E^{bi}$ still has rank one. 
The triad is therefore of type (2,1).

\begin{figure}[bt]
\centerline{
\psfig{figure=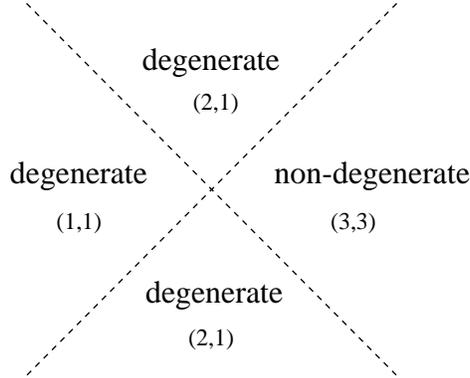,angle=-90,height=5cm}}
\caption{\small Minkowski spacetime spliced onto a degenerate solution
across a pair of intersecting null surfaces. The numbers $(m,n)$ indicate
the rank of $E^a_i$ and $E^a_i E^{bi}$ respectively.}
\label{degenflat}
\end{figure}

In the left hand wedge the triad takes the form
\begin{equation}
E^a_i={\rm diag}(0,0,1),
\end{equation}
which is the type of degenerate triad studied in \cite{1+1}. There it was
found that transverse connection wave excitations are compatible
with this form of triad and propagate at the speed of light.
What happens if such a connection wave
is launched towards the phase boundary from the left hand wedge?
Of course the wave never reaches the phase boundary, which is
also moving at the speed of light. However it is interesting
to ask what will happen when the wave enters
the forward wedge $t>|z|$, where the triad
takes the slightly less degenerate form 
\begin{equation}
E^{x}_{-}=0=E^{y}_{-},\qquad E^{x}_{+}=h(z+t)=-iE^{y}_{+}, 
\qquad E^z_i=(0,0,1).
\end{equation}

Let us examine the constraints and equations
taking the above as an ansatz for $E^a_i$, together with
the ansatz $A_{a}^{3}=0$, $A_z^i=0$. It turns out that this
ansatz is consistent provided that $E^{a}_{+}A_{a}^{+}=0$,
that is, $A_{x}^{+}+iA_{y}^{+}=0$. In particular, if we simply set
$A_{a}^{+}=0$, we have solutions with connection waves propagating
to the right into the forward wedge, without disturbing the
``flat space" form of the triad.

In the next section we will see that one can also splice Minkowski
spacetime across a single null surface.

\section{Covariant approach}

In this section we study the form of 
degenerate phase boundaries by working directly with the spacetime
metric.  We use basically the same simple device as was used
in the canonical approach. The idea (which has occurred to 
others before us) is 
to start from a non-degenerate metric which solves Einstein's equations, 
and then reparametrize one of the coordinates. This reparametrization 
is chosen so that it is {\it not} a diffeomorphism at some particular 
value of the coordinate.  
Adopting the new coordinate, the solution can be smoothly matched to a 
solution to the Ashtekar equations with a degenerate metric at the  
surface where the transformation 
misbehaves.  

What are the regularity conditions we should impose at the phase
boundary?
If the covariant spatial metric is finite, then 
a finite Ashtekar triad density necessarily exists: 
if the metric is diagonalized, $q_{ab}={\rm diag}(q_{xx},q_{yy},q_{zz})$, 
then $E^a_i={\rm diag}
(\sqrt{q_{yy}q_{zz}}\sqrt{q_{zz}q_{xx}}\sqrt{q_{xx}q_{yy}})$.
We shall thus look for solutions where the covariant spatial
metric becomes degenerate, but remains finite, on a surface. 
(Note, however, that given a degenerate Asthtekar triad density 
$E^a_i$, 
the covariant spatial metric is not necessarily finite. For example,
if $E^a_i={\rm diag}(1,1,0)$ then $q_{zz}$ is infinite. Thus we 
exclude from the outset such possibilities from our study in this section.)
In fact we shall restrict attention to cases where the full covariant
spacetime metric is regular. In addition, we must require that 
the lapse with weight minus one be well defined.
Because of eq. (\ref{2}) this is a nontrivial condition which implies
in particular that the scalar lapse must go to zero at the phase boundary.  

\subsection{Flat spacetime}
\label{flat}
The simplest solution that we can consider is the Minkowski space metric 
\begin{equation} 
ds^2 = - dT^2 + dX^2 + dY^2 + dZ^2 \ . 
\end{equation}
Now perform the coordinate transformation
\begin{equation}  
Z = Z(z) \hspace{1cm} \Rightarrow \hspace{1cm} 
ds^2 = - dT^2 + dX^2 + dY^2 + Z'^2dz^2 \,  
\end{equation}
where $Z'=dZ/dz$.
Since we do not insist that the reparametrization should be 
a diffeomorphism we can choose
\begin{equation} 
Z = z^3/3 \ , \ z \geq 0 \hspace{2cm} 
Z = 0 \ , \ z < 0 \ , 
\end{equation}
in which case the metric becomes degenerate at 
$z = 0$. Indeed this example can be generalized to provide metrics 
that are well defined but degenerate anywhere we please, and since the 
spin connection also transforms covariantly it will always be left well 
defined by such a transformation. 

That one can use this device to produce degenerate covariant metrics 
has been noticed before, for instance by Horowitz \cite{Horowitz} who 
also quotes theorems to the effect that if one considers a dense set 
of smooth coordinate transformations then the generic situation is that 
one obtains three dimensional hypersurfaces where the metric has rank 
three and isolated points where it has rank two. 

Now we add the requirement that Ashtekar's lapse should be well 
defined as well. The situation then becomes more restrictive; in the 
example we just considered
\begin{equation} N = \frac{1}{Z'} \ . 
\end{equation}
This would diverge if $Z' \rightarrow 0$, so that the 
example must be rejected. We therefore try something more general:
\begin{equation} T = T(t,z) \hspace{2cm} Z = Z(t,z) \ . 
\end{equation}
To keep things simple we insist that we should have 
$g_{tz} = 0$ 
also after the transformation, which is ensured by the choice
\begin{equation} 
\dot{T} = Z' \hspace{1cm} T' = \dot{Z} 
\hspace{1cm} 
\Rightarrow \hspace{1cm} \ddot{Z} = Z^{\prime \prime} \,
\end{equation}
where a dot and a prime denote differentiation with respect to
$t$ and $z$ respectively.
In words, the simplification implies that $Z(t,z)$ and $T(z,t)$ 
obey the two-dimensional massless  
wave equation, or in other words that the most general 
coordinate transformation still at our disposal expresses $Z$ and $T$ as 
sums of ``left" and ``rightmovers":
\begin{equation} Z = f(z + t) + g(z - t) 
\label{Z}
\end{equation} 
\begin{equation} T = f(z + t) - g(z - t) + \mbox{constant} \ .
\label{T} 
\end{equation}
The line element becomes
\begin{equation} ds^2 = 4f'g'(- dt^2 + dz^2) + 
dX^2 + dY^2 \ . 
\end{equation}
This time Ashtekar's lapse $N$ is unity, and we can 
allow the metric to become degenerate through the choice
\begin{equation} {\rm lim}_{s \rightarrow 0+}\, g^{\prime}(s) = 0 \ . 
\end{equation}
Again the coordinate transformation fails to be a 
diffeomorphism at the surface of degeneracy. The latter is given by 
$z = t$, which when rexpressed in the original coordinates is 
simply the lightlike plane $T = Z + \mbox{constant}$.

This solution can now be joined across the surface of degeneracy to a 
regular solution of Ashtekar's equations which has an everywhere 
degenerate metric. 
How do we know this is possible? One argument goes as follows: 
Extend any timeslice from the nondegenerate region across the 
phase boundary $T = Z + \mbox{constant}$. On this slice use
the Minkowski initial data in the nondegenerate region and
smoothly match to degenerate data on the other side. When this 
initial data is evolved, it will simply yield the Minkowski spacetime
everywhere outside, since the null phase boundary coincides
with the Cauchy horizon for the exterior data. 

Another method is to argue that, 
under a degenerate ``coordinate transformation",
the Ashtekar variables---which can be obtained in a polynomial manner
from the covariant tetrad and connection---remain finite. 
If the lapse and shift also remain finite then it seems reasonable
to suppose that the transformed variables continue to solve 
the Ashtekar equations because the transformation has 
the same form as a diffeomorphism (although it is not a 
diffeomorphism). We have not yet managed to show that this  
is true in general, but it can be explicitly checked
for the coordinate transformation
(\ref{Z}, \ref{T}).
In this case the connection remains zero, the lapse unity, and the shift
zero, and the triad takes the form 
\begin{equation} E^x_i = \left( \begin{array}{c} f' + g' \\ 
i(f' - g') \\ 0 \end{array} \right) \hspace{5mm} 
E^y_i = \left( \begin{array}{c} - i(f' - g') \\ 
f' + g' \\ 0 \end{array} \right) \hspace{5mm} 
E^z_i = \left( \begin{array}{c} 0 \\ 
0 \\ 1 \end{array} \right) \ . \end{equation}
It is easily checked that for {\em any} $f(z+t)$ and $g(z-t)$ 
this is a solution to the Ashtekar equations (\ref{Edot}).
Choosing $f'\ne0$ and $g'(s) = 0 $ for $ s < 0 $ yields 
a solution with a single null degeneracy boundary. To recover the
solution of the previous section with a pair of intersecting
phase boundaries one must choose  $f'=g'$.

\subsection{Spherically symmetric spacetime}
\label{sph}
The general structure of the result just obtained  depends neither 
on the plane symmetry nor on the flatness of Minkowski space.   For
example, it is possible  to perform a coordinate transformation such
that the metric becomes  degenerate on the forward light cone of a
point in Minkowski space.   In this case  the surface of degeneracy
is singular  at the tip of the  cone,   so we can not really  match
the solution smoothly to an everywhere degenerate solution on the  other
side.  

This problem is avoided 
if instead we consider the 
more general spherically symmetric line element
\begin{equation} ds^2 = - F(R)dT^2 + \frac{dR^2}{F(R)} + 
R^2d{\Omega}^2 \ , 
\label{sphline}
\end{equation}
where
\begin{equation} F(R) = 1 - \frac{2m}{R} 
\end{equation}
for the Schwarzschild solution, although the explicit form 
of $F(R)$ will not matter for the argument which follows. If we let 
$R$ and $T$ depend on $t$ and $r$, and then simplify things by the 
requirements
\begin{equation} 
\dot{T} = \frac{R'}{F(R)} \hspace{2cm} 
T' = \frac{\dot{R}}{F(R)} \ , 
\label{simple}
\end{equation}
(with dot and prime denoting differentiation with respect to $t$ and $r$
respectively),
we find that the shift vector vanishes and that the line 
element becomes
\begin{equation} 
ds^2 = \frac{R'^2 - \dot{R}^2}{F(R)}
( - dt^2 + dr^2) + R^2 d{\Omega}^2 \ . 
\label{line}
\end{equation}
Ashtekar's lapse is
\begin{equation} 
N = \frac{1}{R^2\sin{\theta}} \ . 
\end{equation}
(The coordinate singularities at the poles are benign and do not 
concern us here.)
The simplifying requirements (\ref{simple}) imply that 
\begin{equation} 
\ddot{R} - R^{\prime \prime} = 
F^{-1}F_R (\dot{R}^2 - R^{\prime 2}) \ ,
\label{c}
\end{equation}
where $F_R$ denotes $dF/dR$.

Now the spatial metric becomes degenerate only if
\begin{equation} 
R'^2- \dot{R}^2 = 0 \ . 
\label{wave}
\end{equation}
As such a surface of degeneracy is approached, the right hand side 
of (\ref{c}) approaches zero, so the function $R(t,r)$ approaches 
a solution to the two-dimensional massless 
wave equation. Near the surface $R(r,t)$ 
can thus be written in the approximate form
\begin{equation} 
R \approx f(r + t) + g(r - t) \ . 
\end{equation}
Now (\ref{wave}) implies $f'g'\approx 0$, so we have  
either $f^{\prime} = 0$ or $g^{\prime}  = 0$ at the surface of
degeneracy.   This means that the latter is given  by $r = \pm t \ + c$
for some constant $c$. This is a null  hypersurface, as seen from  the form
of the line element (\ref{line}).  For the Schwarzschild  solution
this hypersurface remains regular all the way back to the  white hole
singularity (or to the naked singularity in the negative mass case). 

Let us stress that, previous claims \cite{Bengtsson} notwithstanding, 
the matching to a degenerate metric can take place at any 
initial value of the radial coordinate $R$.

\subsection{Collision with a plane wave}
 
\begin{figure}[tb]
\centerline{
\psfig{figure=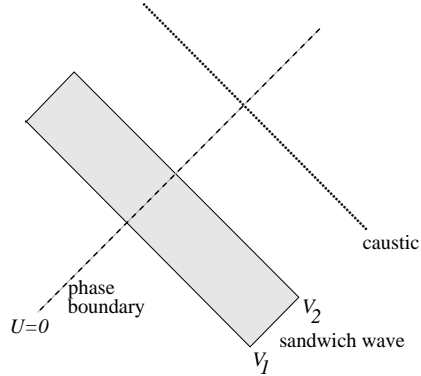,angle=-90,height=5cm}}
\caption{\small  Sandwich plane wave moving leftward collides
with a degenerate phase boundary moving rightward.}
\label{planewave}
\end{figure}

In order to learn something about the stability 
properties of a degenerate region we will verify that a 
degeneracy surface which is a null plane will remain null also if 
it collides with a plane gravitational wave coming from the 
opposite direction. The actual calculation that we will perform 
is a trivial reparametrization of the well known plane wave 
solution, so we begin with a short review of the properties 
of that solution\cite{Bondi,Penrose}. 
The line element has the form 
\begin{equation} ds^2 = - dUdV + F^2dX^2 + G^2dY^2 \ , \end{equation}
where $F$ and $G$ are functions of $V$ only. This metric 
solves Einstein's equations if and only if 
\begin{equation} F G'' + G F'' = 0 \ . 
\label{einstein}
\end{equation}
This is a plane wave moving ``leftwards". Since the 
functions $F$ and $G$ can be chosen at will, subject only to 
the condition (\ref{einstein}), we can consider a ``sandwich wave" which has 
non-vanishing curvature only in some finite interval of the coordinate 
$V$. Then spacetime is flat in front of the wave, and it reverts to 
being flat after the passing of the wave (see Fig. \ref{planewave}). 

We would now like to study a collision between such a plane wave and 
a surface of degeneracy moving ``rightwards". This is simple enough, 
since we can introduce such a surface at $U = 0$ (say) through 
a reparametrization of the coordinate $U=U(u)$. 
For the Hamiltonian decomposition we use the coordinates $T=(U+V)/2$
and $Z=(V-U)/2$, with $T$ the time function and $\partial/\partial T$ the
time flow vector field.
One finds by means of a 
calculation similar to the one we gave in detail for flat spacetime 
that it is possible to perform the reparametrization in such a 
way that Ashtekar's triad $E^a_i$ becomes degenerate at $U = 0$, 
while the lapse, the shift, and the connection remain unaffected.
In a convenient gauge, the self-dual spin connection has the non-vanishing
components
\begin{equation}
A_{X1}=iA_{X2}=F'\qquad A_{Y2}=-iA_{Y1}=G'.
\end{equation}
The only non-zero components of the self-dual curvature are
\begin{equation}
F_{VX1}=iF_{VX2}=F'' \qquad F_{VY2}=-iF_{VY1}=G''.
\end{equation}
These are untouched by the 
reparametrization $U=U(u)$.

In the degenerate region we must have a degenerate solution to the
Ashtekar equations. The form this solution takes is the following.
Define the new coordinates $(t,z)$ via
\begin{equation} U = T - Z = - 2g(z-t) \hspace{1cm} 
V = T + Z = t + z \ , \end{equation}
and introduce the SO(3) triad
\begin{equation} U_i = \left( \begin{array}{c} 1 \\ -i \\ 0 
\end{array}\right) \hspace{5mm} 
V_i = \left( \begin{array}{c} 1 \\ i \\ 0 
\end{array}\right) \hspace{5mm} 
W_i = \left( \begin{array}{c} 0 \\ 0 \\ 1 
\end{array}\right) \end{equation}
($U_i$ and $V_i$ are null vectors). Then the transformed 
Ashtekar variables for the plane wave take the form
\begin{equation} N = \frac{1}{FG} \hspace{2cm} N^a = 
{\Lambda}_i = 0 \end{equation}
\begin{equation} E^x_i = \frac{1}{2}GV_i + g'GU_i 
\hspace{1cm} E^y_i = - \frac{i}{2}FV_i + ig'FU_i \hspace{1cm} 
E^z_i = FGW_i \label{E} \end{equation}
\begin{equation} A_{xi} = F'U_i \hspace{1cm} A_{yi} = iG'U_i 
\end{equation}
\begin{equation} F_{yz\, i} = - iG''U_i \hspace{1cm} F_{zx\, i}= 
F''U_i \ . \end{equation}

All these quantities are well defined when $g' = 0$, even if
$g'(u)=0$ for all $u$ greater than some $u_0$. As discussed in 
section \ref{flat}, it seems likely that in general a degenerate
Ashtekar configuration obtained in this manner will automatically
be a solution if the original nondegenerate configuration
was a solution and the lapse is finite. In this particular case
it is straightforward to check explicitly that all the constraints
and equations of motion are satisfied. Thus, after colliding with 
the degnerate phase boundary, the plane wave is converted into
a degenerate plane wave with the same connection and curvature.
The triad on the other hand has rank two, while $E^a_iE^{bi}$ has rank
one, as can be seen from Eqn (\ref{E}). This is therefore a degenerate
solution of type (2,1). It is {\it not} one of the connection
waves found in \cite{1+1}, since in that paper only type (1,1)
solutions were studied.
 
There are two salient features here: The connection and curvature of
the wave continue unaffected into the degenerate 
region (they are unaffected by the reparametrization), 
and the degeneracy surface remains null and nonsingular 
as viewed from the nondegenerate side.  
It should also be said that there is a 
global difficulty with this degenerate plane wave solution. 
The surfaces of constant $U$ are focused by the wave and 
eventually develop caustic singularities at some value of $V$.
This is a coordinate singularity related to the fact that
the plane wave spacetimes do not admit Cauchy surfaces\cite{Penrose}. 
These caustics present a problem for us since it
is not at the moment clear how to make sense of a surface 
of degeneracy which is not smooth.

\section{Is the phase boundary always null?}
\label{null?}

The examples studied above suggest that the phase boundary
is always null. We have not so far been able
to prove this, however we offer in this section some remarks
that may be useful in addressing this problem. In particular,
we argue that the equations of motion are at least in part 
necessary, and we conjecture why they might be sufficient.

First of all, what exactly do we mean by the phase boundary 
being null? Since the metric on the nondegenerate side 
is perfectly regular all the way
up to the phase boundary, it is meaningful to say the boundary
is null ``as viewed from the nondegenerate side". More precisely,
suppose the surface on which the metric becomes degenerate is 
given by $f=0$, where $f$ is a smooth function with 
$\nabla_a f|_{f=0}\ne0$ 
with respect to local charts in
which the metric does not degenerate as $f\rightarrow0$.
We consider the surface to be null if 
$g^{ab}\nabla_a f\nabla_b f\rightarrow 0$ as the surface is 
approached.

The regularity of the Ashtekar variables at the surface 
seems to be necessary
for the null conjecture to have any chance of being true. 
Counterexamples serve to illustrate this.
First, consider the reparametrization $Z=z^3/3$ of Minkowski space
discussed in section \ref{flat}. There the degeneracy
surface $Z=0$ is clearly {\it timelike}. However, when the 
coordinate $z$ is adopted,
the Ashtekar lapse density diverges, so this configuration
does not qualify as a regular Ashtekar solution. As a second example
consider the FRW line element 
$ds^2=-dt^2 + t(dx^2+dy^2+dz^2)$. For this line element, the 
spatial metric becomes 
degenerate on the {\it spacelike} surface $t\rightarrow 0^+$.
Here also the lapse density diverges so, again, one does not
have a regular Ashtekar solution at the boundary.

The regularity of the lapse density, although necessary, 
is not sufficient to ensure nullness of the phase boundary.
It seems that, at least to some degree, the equations of motion 
are  required. In the spherically symmetric case, 
analyzed in section \ref{sph}, it was found that the phase 
boundary is always null.
No use was made of the particular
form of the function $F(R)$ in the line element (\ref{sphline}),
and of course for arbitrary $F(R)$ one does not have a solution
to the Einstein equation. On the other hand, one might argue that
{\it part} of the equations of motion has been used in restricting 
to the case $g_{tt}g_{rr}=-1$, and perhaps this part of the
equations of motion is enough to imply nullness of the phase boundary
in this very symmetrical case.
To further explore this question here we will generalize the previous
analysis to allow for independent coefficients of $dt^2$ and $dr^2$ 
in the spherically symmetric line element. We find that the
boundary is {\it not} null in the general (non-solution) case.
 
Consider a line element of the form
\begin{equation} ds^2 = - F(R)dT^2 + G(R) {dR^2}  + 
R^2d{\Omega}^2, 
\label{FG}
\end{equation}
where 
\begin{equation}
H:=FG 
\end{equation}
is not necessarily constant. The case $H(R)=1$ corresponds to the
previously considered situation. Now let $R$ and $T$ depend 
on $t$ and $r$, and make the simplifying requirements
\begin{equation} 
\dot{T} = R'/F\hspace{2cm} 
 {T}' =  G{\dot{R}} 
\label{simple2}
\end{equation}
so that the metric will remain diagonal. 
The line element then becomes
\begin{equation} 
ds^2 =  F^{-1}({R}'^2 - H\dot{R}^2)
( - dt^2 + Hdr^2) + R^2 d{\Omega}^2,
\label{line2}
\end{equation}
and Ashtekar's lapse density is
\begin{equation} 
N = \frac{1}{H^{1/2}R^2\sin{\theta}} \ . 
\end{equation}
The simplifying requirements (\ref{simple2}) imply that 
\begin{equation} 
H\ddot{R} - R^{\prime \prime} = 
F^{-1} {F_R} (H\dot{R}^2 - R^{\prime 2})-H_R \dot{R}^2 \ , 
\label{ddotR}
\end{equation}
where the subscript $R$ denotes differentiation with respect to $R$.

Now the spatial metric becomes degenerate  at the 
surface 
\begin{equation} 
f:=R^{\prime 2} - H\dot{R}^2 = 0 \ . 
\label{wave2}
\end{equation}
To see whether this surface is null we examine the normal 
vector\footnote{It is important here that $\nabla_a f$ is a non-zero
vector where $f=0$, i.e. that $\dot{f}$ and $f'$ are nonzero
there. Although we have not proven that they are nonzero, there 
appears to be no reason why they should vanish.}
\begin{equation}
g^{ab}\nabla_a f\nabla_b f=F(Hf)^{-1}[f'^2-H\dot{f}^2]
\label{normal}
\end{equation}
in the limit that the surface $f=0$ is approached.
Factorizing the last term, we are led to evaluate
\begin{equation}
f'\pm H^{1/2}\dot{f}=
(2R''\pm H^{1/2}\dot{R}'-H_R\dot{R}^2)
(R'\mp H^{1/2}\dot{R})
\pm 2H^{1/2}\dot{R}(F^{-1}F_R f+H_R\dot{R}^2),
\label{factor}
\end{equation}
where (\ref{ddotR}) has been used to eliminate the $\ddot{R}$ term.
If $H$ is constant, then the last term of (\ref{factor})
vanishes as the $f=0$ surface is approached,
and we have
\begin{equation}
(f'+H^{1/2}\dot{f})(f'-H^{1/2}\dot{f})
\propto 
(R'+H^{1/2}\dot{R})(R'- H^{1/2}\dot{R})=f=0,
\end{equation}
so the right hand side of (\ref{normal}) vanishes, so
the $f=0$ surface is null. If $H$ is {\it not} constant,
however, the surface appears {\it not} to be null.

We conclude from this analysis that the phase boundary  
generated by the degenerate coordinate transformation is 
null only if $H:=FG$ is constant in the line element (\ref{FG}).
Thus at least part of the Einstein equation is required.
We suspect that the nullness of the phase boundary is related
to the fact that the characteristic surfaces of the (nondegenerate)
Einstein equation are null. Since the Einstein equation locally 
preserves the rank of the metric, it seems likely that the rank can 
only change across a characteristic surface. We conjecture that this 
is the underlying reason for the nullness
of the phase boundary in the cases examined so far.
 
\section{Discussion}

We have obtained in this paper  
examples of solutions of Ashtekar's equations in which degenerate 
and non-degenerate metrics coexist, separated by a ``phase boundary".
These ``geometries" differ from the kind of degenerate solutions that 
occur in other degenerate extensions of general relativity.  

We think that our examples provide a reasonable amount of support 
for our conjecture that the boundary of the degenerate region is 
always null, as viewed from the non-degenerate side. 
This appears to make some sense, since the characteristic 
surfaces on the nondegenerate side are null. 
It also ties in well with Matschull's observation 
\cite{Matschull} that the specific way in which Ashtekar's 
variables allow the metric to become degenerate is such that 
a local causal structure of spacetime is 
preserved in the form of a partly collapsed lightcone. In the examples
we have studied, 
the phase boundary is also null, in Matschull's sense,
as viewed from the degenerate side, so the phase boundary 
propagates causally in that sense too.

All of our examples have the feature that the phase boundary
is a smooth null hypersurface (except at the focal surface in 
the plane wave spacetime). It is not at all clear whether 
degenerate and nondegenerate solutions can be matched across
null surfaces that are {\it not} smooth.
Since caustics on null
surfaces are generic, the scope of the investigation should
be expanded to determine what happens at caustics. This could
even be investigated in the flat space example, taking as the 
initial phase boundary a dimpled surface whose null normal congruence
develops a caustic.

In this paper we have restricted attention to the vacuum case only.
For a general matter stress-energy tensor the static, spherically
symmetric solution does not in general have constant $H$,
so the results of section \ref{null?} seem to suggest that even if 
the Einstein equation is satisfied the phase boundary may fail to be null.
However, if matter is included, one must also 
require that the canonical matter variables in the Ashtekar
formulation are regular as the surface of degeneracy is approached.
If they are not, then the phase boundary is not allowed in a 
regular solution. It would be interesting to check this in some 
examples. The Reissner-Nordstrom solution has constant $H$, so 
one needs to look elsewhere. For instance, a self-interacting,
self-gravitating scalar field configuration could provide 
an example, or a stellar interior with perfect fluid matter.
Another interesting example would be an electrovac plane 
wave. 
Scalar and electromagnetic matter fields\cite{AshtRomaTate} 
and perfect fluids\cite{BombTorr} 
have all been incorporated into the Ashtekar formulation already, 
so the groundwork has been laid.  

A question we have ignored, but which should be answered
if the degenerate extension of the theory is to be 
taken seriously, is whether the initial value problem is
in fact well-posed. The answer probably depends on the
way in which the metric is allowed to become degenerate.
For example, when it is the covariant metric that is
allowed to be degenerate, it was shown by Horowitz\cite{Horowitz}
that the topology of spacetime is not determined by initial
data since the topology can change unpredictably. In the 
degenerate Plebanski formulation, Reisenberger\cite{Reisenberger}
showed  
that initial data does not determine a solution for the
fields even in fixed toplogy. The status of the initial value problem
in the degenerate Ashtekar theory is not known (as far as we know),
but the existence of the (degenerate) causal structure\cite{Matschull}
suggests that the dynamics may indeed be determined by data
on a spacelike surface. In particular, the causal structure
implies the existence of a nowhere vanishing ``timelike"
vector field which, together with a causality condition ruling out
closed timelike curves,
may be enough to yield the result (proved by Geroch\cite{Gero}
for nondegenerate Lorentzian metrics) that topology change in a spatially
compact universe is impossible.  
 
Lest we get carried away extolling the virtues of the 
degenerate Ashtekar theory, it is worth restating the fact, 
mentioned in the introduction, that this theory seems
to admit negative energy configurations\cite{Vara}, whereas
other degenerate extensions of general relativity do not\cite{Sam}. 
It remains very much an open question which, if any, degenerate extension 
is physically reasonable, not to mention which, if any, is actually correct.
 
\section{Acknowledgments}

We thank Claes Uggla for discussions at an early stage of 
this work. The work of
TJ was supported in part by NSF grant PHY94-13253, the Institute
for Theoretical Physics at the University of Utrecht, the General Research
Board of the University of Maryland, and the Erwin Schr\"odinger Institute. 
IB was partially supported by the NFR.

\end{document}